# Systematic Study of Electronic Phases, Band Gaps and Band Overlaps of Bismuth Antimony Nanowires

Shuang Tang*[1], Mildred S. Dresselhaus†[2,3]

[1]Department of Materials Science and Engineering, Massachusetts Institute of Technology, Cambridge, MA, 02139-4037, USA; *tangs@mit.edu

[2]Department of Electrical Engineering and Computer Science, Massachusetts Institute of Technology Cambridge, MA, 02139-4037, USA; †millie@mgm.mit.edu

[3]Department of Physics, Massachusetts Institute of Technology Cambridge, MA, 02139-4037, USA;

**Abstract:** We have developed an iterative one dimensional model to study the narrow band-gap and the associated non-parabolic dispersion relations for bismuth antimony nanowires. An analytical approximation has also been developed. Based on the general model, we have developed, we have calculated and analyzed the electronic phase diagrams and the band-gap/band-overlap map for bismuth antimony nanowires, as a function of stoichiometry, growth orientation, and wire width.

## I. INTRODUCTION

The bismuth antimony ($Bi_{1-x}Sb_x$) alloys material are considered to be one of the best materials for low temperature thermoelectrics, supercooling, millivolt electronics and infrared applications. A notable number of interesting properties have been observed in bulk bismuth materials, such as non-parabolic dispersions and abnormal magneto-resistance, [1] the ultra-high mobility of carriers, [2] and the high anisotropy.[3] In 1993, Hicks et al. suggested that thermoelectric materials could have enhanced figure of merit if the materials were synthesized in the form of low dimensional systems and nano-systems. [4] Since then, much more focus has been given to bismuth antimony as related to nanoscience and nanotechnology. Lin et al. have synthesized and studied some electronic properties of $Bi_{1-x}Sb_x$ nanowires of diameters (d=40, 45 and 60 nm) and Sb composition (x=0, 0.05, 0.10 and 0.15), which are oriented along the (012) crystalline axis. [5] $Bi_{1-x}Sb_x$ nanowires with their wire axis oriented in the trigonal direction have been studied by Rabin et al, [6] for the composition range of (x=0~0.30) and the nanowire diameter range (d=10~100 nm), below the temperature of 77 K, where the authors suggested that the thermoelectric performance can be optimized by aligning the carrier pockets. $Bi_{1-x}Sb_x$ wires of larger diameters on the order of microns have been studied for their strain effect in the wire direction on their electrical resistivity by Nikolaeva et al. [7] Tang and Dresselhaus have given systematic guidance on the electronic band structure of $Bi_{1-x}Sb_x$ thin films as a function of growth orientation, film thickness, stoichiometry and temperature. [8,9] However, there are no corresponding systematic studies of $Bi_{1-x}Sb_x$ nanowires, and researchers, especially experimentalists, are very eager for global guidance on $Bi_{1-x}Sb_x$ nanowires as a function of growth orientation, wire diameter, stoichiometry, temperature, etc. The present paper aims to provide such guidance on the electronic phases and band gaps or band overlaps of $Bi_{1-x}Sb_x$ nanowires to stimulate the synthesis of such nanowires for different applications.

In the current paper, we first develop a model for the mini-band gap and the related non-parabolic dispersion relations at the *L* point of bismuth antimony in one dimension. In particular we use an the iterative one dimensional two band model, and we have here developed an analytical approximation for this model. Thereafter, we study the band edges and electronic phases as a function of growth orientation, wire diameter and stoichiometry, including the semimetal phases, the indirect semiconductor phases and the direct semiconductor phases. The band overlap of the semimetal phases, and the band gap of the semiconductor phases are then studied as well. The aim of this paper is to: 1) develop a one dimensional non-parabolic band model, and 2) provide a guide for the synthesis of $Bi_{1-x}Sb_x$ nanowires with various crystallographic orientations that could be used for different applications.

We first review the crystal and electronic band structure of bulk materials of $Bi_{1-x}Sb_x$. The bulk materials of bismuth, antimony and their alloys have the same $R\bar{3}m$ symmetry with a rhombohedral lattice structure. There are two atoms per unit cell, as shown in Fig. 1a. The trigonal axis with a three-fold symmetry, the binary axis with a two-fold symmetry, and the bisectrix axis form a natural Cartesian coordinate frame as shown in Fig. 1(a). A mirror plane is formed by the trigonal axis and the bisectrix axis, which is perpendicular to the binary axis. In the first Brillouin zone, the band edges are located at the *T* point, the three *L* points, and the six *H* points, [10,11] as shown in Fig. 1b. In the range of (x=0~0.30), the bottom of the conduction band is always located at the *L* points, while the top of the valence band can be located at the *T* point, the *L* points or the *H* points depending on the Sb concentration. The relation between Sb composition (x=0~0.30) and the relative positions in energy of the band edges at different locations in the first Brillouin zone changes with temperature, pressure, external magnetic field, strain, etc. Figure 1(c) shows the case where the temperature is below 77 K, and the pressure is at 1 atm with no external magnetic field.[12] Other cases have similar relations, but different values for the critical Sb concentration for the various phase transitions.[10,13] The conditions of the sample and measurements using different experimental techniques may also results in different values of x at certain phase transition points.[14] In this paper, we will just concentrate on the case of Fig. 1(c), where the band structure does not change notably with temperature. Other cases can be studied in a similar way.

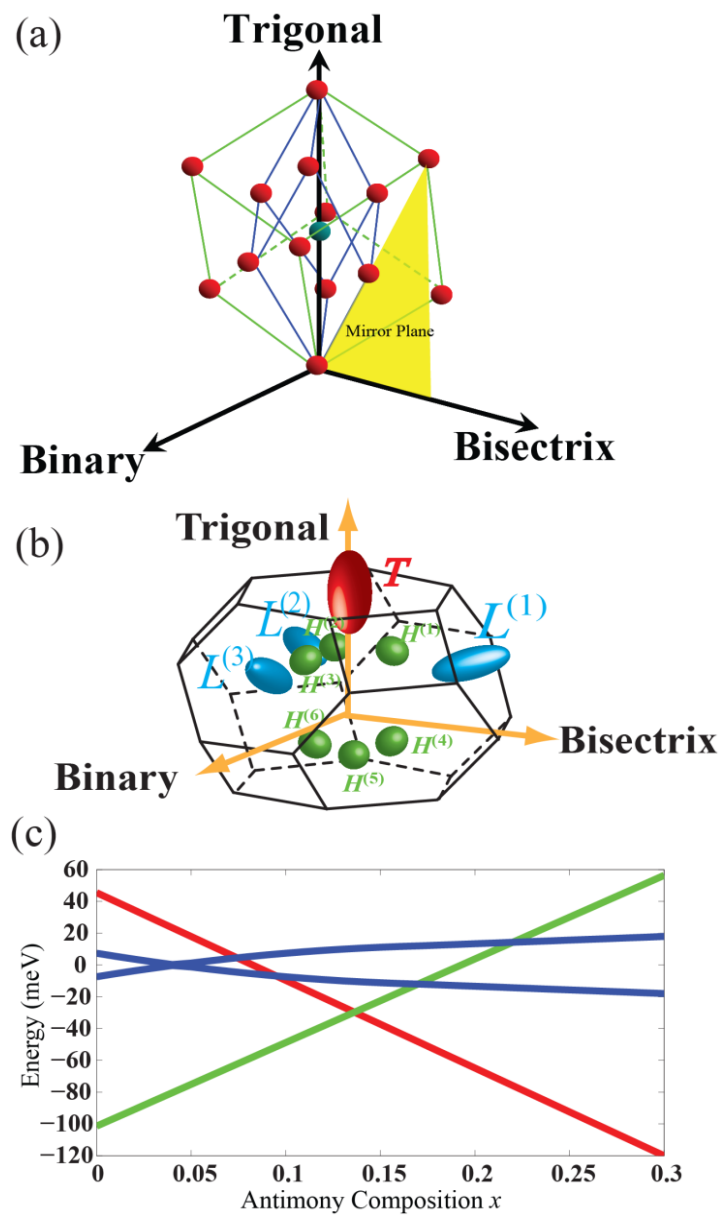

Figure 1: Atomic and band structure of bulk bismuth antimony. (a) The atomic unit cell of bismuth

antimony. The two atoms in each unit cell are marked by green and red. The trigonal, binary and bisectrix axes form a natural three-dimensional Cartesian coordinates framework. (b) The different carrier pockets at the *T* point for holes, the three *L* points for holes and electrons, and the six *H* points for holes. (c) How the band edges at the *T* point, the three *L* points and the six *H* points change as a function of Sb composition x in bulk $Bi_{1-x}Sb_x$.

## II. METHODOLOGY

The quantum confinement effect of nanowires may change the symmetry properties of the carrier pockets, the positions in energy of the band edges, and possibly the shape of the dispersion relations at the *T* point, the *L* points and the *H* points. There are two types of quantum confinement effects on the band edges. One is a trivial quantum confinement effect that does not change the inverse-effective-mass tensor or the shape of the dispersion; the other is a non-trivial quantum confinement effect that does change the inverse-effective-mass tensor and the shape of the dispersion. For the *T* point and the *H* points, the dispersions of the band edges are parabolic, and the quantum confinement effect is trivial. The valence band edge at the *T* point will decrease in energy by $h^2/8 \cdot (d_1^{-2} \cdot \alpha_{\parallel,11}^{[T,\text{Wire}]} + d_2^{-2} \cdot \alpha_{\parallel,22}^{[T,\text{Wire}]})$, where $d_1$ and $d_2$ are the cross-sectional widths of a rectangular nanowire, and $\alpha_{\parallel,11}^{[T,\text{Wire}]}$ and $\alpha_{\parallel,22}^{[T,\text{Wire}]}$ are the corresponding components of the cross-sectional inverse-effective-mass tensor.[15] For a square nanowire, the width is $d = d_1 = d_2$, and the expression of energy decrease at the *T* point is reduced to $h^2/8d^2 \cdot (\alpha_{\parallel,11}^{[T,\text{Wire}]} + \alpha_{\parallel,22}^{[T,\text{Wire}]})$. Similarly, the valence band edge at an *H* point will decrease in energy by $h^2/8 \cdot (d_1^{-2} \cdot \alpha_{\parallel,11}^{[H,\text{Wire}]} + d_2^{-2} \cdot \alpha_{\parallel,22}^{[H,\text{Wire}]})$ for a rectangle nanowire, and by $h^2/8d^2 \cdot (\alpha_{\parallel,11}^{[H,\text{Wire}]} + \alpha_{\parallel,22}^{[H,\text{Wire}]})$ for a square nanowire. Furthermore, we can still assume that

$\boldsymbol{\alpha}_{\parallel}^{[T,\text{Wire}]} = \boldsymbol{\alpha}_{\parallel}^{[T,\text{Bulk}]}$ and $\boldsymbol{\alpha}_{\parallel}^{[H,\text{Wire}]} = \boldsymbol{\alpha}_{\parallel}^{[H,\text{Bulk}]}$ as have been validated in previous reports [5,6,8,9].

For the *L* points, the dispersions of the band edges are non-parabolic, and the quantum confinement effect is non-trivial, so that the traditional square-well model is not accurate any more. There coexists an electron pocket and a hole pocket at each *L* point. Thus, the conduction band edge and the valence band edge for the *L* point are very close to each other in energy, and are strongly coupled, which results in the non-parabolicity of the dispersion for the *L*-point electrons and holes or possibly even linearity if the two bands are touching. Meanwhile, the shape of the dispersion is also correlated with the magnitude of the narrow band gap. On the one hand, quantum confinement in the cross-sectional plane will change the narrow band gap, which is associated with the inverse-effective-mass tensor. On the other hand, the inverse-effective-mass tensor of $Bi_{1-x}Sb_x$ nanowire is changed by the change of the narrow band gap, which is also different from the inverse-effective-mass tensor of a bulk $Bi_{1-x}Sb_x$ with the same Sb composition. Therefore, the puzzle is that neither the narrow band gap nor the inverse effective-mass tensor is known at this stage for the nanowires case.

Historically, the relation between the narrow band gap and the dispersion for bulk bismuth is described by a two-band model,[16]

$$\mathbf{p} \cdot \boldsymbol{\alpha}^{[L,\text{Bulk, Bi}]} \cdot \mathbf{p} = E(\mathbf{k})(1 + \frac{E(\mathbf{k})}{E_g^{[L,\text{Bulk, Bi}]}}), \quad (1)$$

where $\boldsymbol{\alpha}^{[L,\text{Bulk}]}$ is the inverse-effective-mass tensor for the L point carrier-pocket of bulk bismuth, and it is assumed that $\boldsymbol{\alpha}^{[L,\text{Bulk}]}$ for both the conduction band edge and the valence band edge are the same due to the strong interaction between these two bands and this approximation has been shown to be valid for bulk bismuth samples. The relation between the band gap and the inverse-effective-mass tensor is seen more clearly in the form of the second derivative of Eq. (1),[17]

$$\boldsymbol{\alpha}^{[L,\text{Bulk, Bi}]} = \frac{2}{\hbar^2}\frac{\partial^2 E(\mathbf{k})}{\partial \mathbf{k}^2} = \frac{1}{m_0}\mathbf{I} \pm \frac{1}{m_0^2}\frac{2}{E_g^{[L,\text{Bulk, Bi}]}}\mathbf{p}^2, \quad (2)$$

where $m_0$ is the mass of free electron and $\mathbf{I}$ is the identity matrix. It is assumed that the form of Eq. (2) holds also for $\text{Bi}_{1-x}\text{Sb}_x$ in the range of x=0~0.30, though the band gap and the effective mass tensor may change as a function of Sb composition x. Thus, we further have, [5,18]

$$\boldsymbol{\alpha}^{[L,\text{Bulk, Bi}_{1-x}\text{Sb}_x]} = \frac{E_g^{[L,\text{Bulk, Bi}_{1-x}\text{Sb}_x]}}{E_g^{[L,\text{Bulk, Bi}]}}(\boldsymbol{\alpha}^{[L,\text{Bulk, Bi}]} - \frac{1}{m_0}\mathbf{I}) + \frac{1}{m_0}\mathbf{I}, \quad (3)$$

which is consistent with the experimental results carried out by Mendez et al, where a simpler version of Eq. (3) was adopted [19] and is given by,

$$\boldsymbol{\alpha}^{[L,\text{Bulk, Bi}_{1-x}\text{Sb}_x]} = \frac{E_g^{[L,\text{Bulk, Bi}_{1-x}\text{Sb}_x]}}{E_g^{[L,\text{Bulk, Bi}]}}\boldsymbol{\alpha}^{[L,\text{Bulk, Bi}]}. \quad (4)$$

The theoretical validation between from Eq. (3) to Eq. (4) is discussed in Ref. [8], and these experiences can each be extended to connect the bulk materials and nanowire materials, i.e.

$$\boldsymbol{\alpha}^{[L,\text{Wire, Bi}_{1-x}\text{Sb}_x]} = \frac{E_g^{[L,\text{Wire, Bi}_{1-x}\text{Sb}_x]}}{E_g^{[L,\text{Bulk, Bi}]}}(\boldsymbol{\alpha}^{[L,\text{Bulk, Bi}]} - \frac{1}{m_0}\mathbf{I}) + \frac{1}{m_0}\mathbf{I}, \quad (5)$$

and

$$\boldsymbol{\alpha}^{[L,\text{Wire, Bi}_{1-x}\text{Sb}_x]} = \frac{E_g^{[L,\text{Wire, Bi}_{1-x}\text{Sb}_x]}}{E_g^{[L,\text{Bulk, Bi}]}}\boldsymbol{\alpha}^{[L,\text{Bulk, Bi}]}. \quad (6)$$

Now we have two approaches to solving the dispersion relation and for finding the narrow band gap of the $L$-point band edges. One is the iterative way. We set $E_g^{(0)} = E_g^{[L,\text{Bulk, Bi}]}$ and $\boldsymbol{\alpha}^{(0)} = \boldsymbol{\alpha}^{[L,\text{Bulk, Bi}]}$, and we repeatedly carry out the iteration steps of

$$E_g^{(n+1)} = E_g^{[n]} + 2\cdot\frac{h^2}{8d^2}\cdot\text{trace}(\boldsymbol{\alpha}_\parallel^{(n)}) \quad (7)$$

and

$$\boldsymbol{\alpha}^{(n+1)} = \frac{E_g^{(n)}}{E_g^{(n+1)}}(\boldsymbol{\alpha}^{(n)} - \frac{1}{m_0}\mathbf{I}) + \frac{1}{m_0}\mathbf{I} \quad (8)$$

until convergence, where $\boldsymbol{\alpha}_{\parallel}^{(n)}$ is the cross-sectional sub-tensor of $\boldsymbol{\alpha}^{(n)}$.

The other approach is to use the simpler Eq. (6), and derive a solution in an analytical form, by solving Eq. (6) and

$$E_g^{[L, \text{Wire}, \text{Bi}_{1-x}\text{Sb}_x]} = E_g^{[L, \text{Bulk}, \text{Bi}]} + 2 \cdot \frac{h^2}{8d^2} \cdot \text{trace}(\boldsymbol{\alpha}_{\parallel}^{[L, \text{Wire}, \text{Bi}_{1-x}\text{Sb}_x]}) \quad (9).$$

Thus, the L-point narrow band gap of the nanowire can be solved to be,

$$E_g^{[L, \text{Wire}, \text{Bi}_{1-x}\text{Sb}_x]} = \frac{E_g^{[L, \text{Bulk}, \text{Bi}_{1-x}\text{Sb}_x]} + \sqrt{(E_g^{[L, \text{Bulk}, \text{Bi}_{1-x}\text{Sb}_x]})^2 + \frac{h^2}{d^2} \cdot \text{trace}(\boldsymbol{\alpha}_{\parallel}^{[L, \text{Bulk}, \text{Bi}]}) \cdot E_g^{[L, \text{Bulk}, \text{Bi}]}}}{2}.$$

(10)

## III. RESULTS AND DISCUSSION

Figure 2 illustrates the electronic phase diagrams and band gap/overlap of $\text{Bi}_{1-x}\text{Sb}_x$ nanowires with d=100 nm, as a function of growth orientation and stoichiometry. The illustrations are made for the growth orientation within the binary plane, the trigonal plane and the bisectrix plane, in Fig. 2 (a), (b) and (c), respectively. The semimetal (SM) phase regions, indirect semiconductor (ISC) phase regions and the direct semiconductor (DSC) phase regions are marked for each orientation. The Sb composition x is denoted by the radius of the circles in Fig. 2. With 100 nm wire width, it can be seen that the electronic phase starts from a semimetal, where the top of the valence band edge is located at the *T* point, at x=0. As the Sb composition increases, phase changes occur. Explicitly, the electronic phase changes from a semimetal to an indirect semiconductor with the top of the valence band edge located at the *T* point, too. At around x=0.15, the top of the valence band edge become located at an *L* point, and the electronic phase becomes a direct semiconductor. For yet higher Sb concentration x, the top of the valence band edge is shifted to an *H* point, and the band gap of the semiconductor phase becomes indirect again. When x is further increased, the electronic phase finally is changed back to a semimetal, only with the top of the valence band edge located at an *H* point.

This is seen more clearly in the maps of the band gap/overlap as a function of growth orientation and Sb composition, in the corresponding subfigures below, i.e. in Fig. 2 (d), (e) and (f), for the binary plane, the trigonal plane and the bisectrix plane, respectively. A negative value stands for a band overlap, while a positive value stands for a band gap. A zero value denotes the gapless state. At Sb composition x=0, the band overlap exhibits a negative value corresponding to the semimetal phase. As x is increased, the magnitude of the band overlap starts to decrease, and then becomes zero, beyond which the band overlap disappears and the nanowire exhibits a band gap with a positive value, and the band gap increases with increasing Sb composition x. At around x=0.15, the band gap starts to decrease with increasing Sb composition x, until the band gap reaches zero. At yet higher Sb composition x, a band overlap with a negative value appears again, indicating the onset of a semimetallic phase at the relatively Sb rich side of the phase diagram (see Fig. 1(c)).

The anisotropic properties of bulk bismuth antimony phases are reflected in the symmetry properties of the electronic phase diagrams and the band gap/overlap diagrams, when the growth orientation lies in different crystallographic planes, or normal to different crystallographic directions. The diagrams of Fig. 2(a), (b) and (c) have similar profiles, but are actually different in their detailed shapes at each phase boundary, if they are examined more closely. The binary crystalline plane has inversion symmetry but not mirror symmetry. The inversion symmetry is reflected in Fig. 2 (a) and (d). However, at this specific wire width, the contrast associated with the anisotropy is not strong, and the non-existence of mirror symmetry is not obvious. We discuss below how quantum confinement helps to enhance the contrast of the anisotropy in the electronic phase diagrams and the band gap/overlap diagrams, and also how quantum confinement causes the mirror symmetry to disappear for the diagrams associated with the binary crystalline plane. When all six *H*-point hole-pockets, all six *L*-point half-hole-pockets and all six *L*-point half-electron-pockets are projected onto the trigoanl plane, a six-fold symmetry is formed, which is a higher symmetry than the three-fold symmetry of the trigonal axis in the bulk materials. This is reflected in the electronic phase diagram of Fig. 2 (b), and in the band gap/overlap phase diagram of Fig. 2 (e). In Fig. 2(b), the shape of each phase boundary is actually a mixture of a circle and hexagon. To see this, it requires a very careful reading of the diagram, and it might be not obvious. However, this circle and hexagon are much more clear in the band gap/overlap diagram in Fig. 2 (e), where the shapes of six-petal flowers can best be seen at the edge of the inner yellow DSC (direct semiconductor phase) region. The diagrams of the bisectrix crystalline plane have both mirror symmetry about the binary axis, and mirror symmetry about the trigonal axis, which are companied by inversion symmetry. These symmetry properties are seen in Fig. 2 (c), and more obviously in Fig. 2(f).

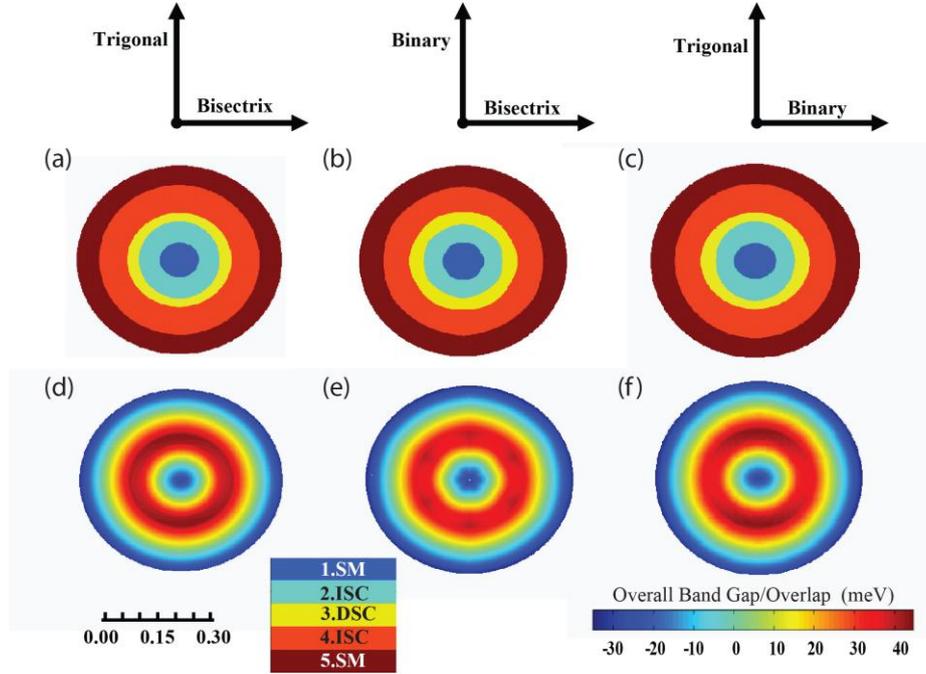

Figure 2: The electronic phase diagrams (a)-(c), and the band gap/overlap diagrams (d)-(f) of $Bi_{1-x}Sb_x$ nanowires of 100 nm wire width, as shown for various wire growth orientations and Sb composition x. The illustrations are made for the binary (a) and (d), trigonal (b) and (e), as well as and bisectrix (c) and (f) crystalline planes. In each diagram, the direction stands for growth orientation, and the length of radius for a point stands for the Sb composition x, which lines up with the origin of each circularly shaped diagram, so that the binary direction in (a) for example denotes the direction normal to the trigonal and bisectrix directions shown for the square sample faces. At the origin, the value of x is 0, and at the outmost point, the value of x is 0.30, as scaled in the legend. The regions semimetal phase (SM), indirect semiconductor phase (ISC) and direct semiconductor phase (DSC) are marked out in (a)-(c) and in the legend. In the legend of the phase regions, the top of the valence band edge is located at the $T$ point for the upper regions of the semimetal (Region 1) and indirect semiconductor phases (Region 2), and at the $H$ points for the lower regions of the indirect semiconductor phase (Region 4) and semimetal phase (Region 5). Both the top of the valence band edge and the bottom of the conduction band edge are located at the *L* points for the direct semiconductor phase (Region 3). In the diagrams of the band gap/overlap, a positive value stands for a band gap, while a negative value stands for a band overlap. A zero value stands for a gapless state.

Now we illustrate the electronic phase diagrams and band gap/overlap diagrams of $Bi_{1-x}Sb_x$ nanowires with a much stronger quantum confinement effect, occurring in nanowires with a small width, explicitly for nanowire with a width of 10 nm. The small width nanowires show how the quantum confinement effect influences the symmetry properties and the electronic phases of the nanowires comparatively. The electronic phase diagrams and band gap/overlap of $Bi_{1-x}Sb_x$ nanowires with d=10 nm, as a function of growth orientation and stoichiometry, are illustrated in Fig. 3. The changes in the electronic phase diagrams are more obvious in Fig. 3 than in Fig. 2. First, the direct semiconductor phase regions have disappeared in all of the three cases (a), (b) and (c). The semimetal phase region (dark blue) where the top of the valence band edge is located at the *T* point has significantly shrunk to a tiny size in both Fig. 3(a) and Fig. 3(c), and has disappeared in Fig. 3(b). The semimetal phase region where the top of the valence band edge is located at an *H* point (light blue) has shrunk as well, in all the three cases, but still is present. The dominant phase regions become the indirect semiconductor phases arrange, which have both expanded remarkably in Fig. 3 (a), (b) and (c). Such information is very important for the design of electronic devices using $Bi_{1-x}Sb_x$ nanowires. The much stronger quantum confinement effect in 10 nm wide nanowires makes the contrast for the anisotropy of all the diagrams much more obvious. The existence of inversion symmetry and the absense of mirror symmetry is shown clearly in Fig. 3 (a) and (d). In Fig. 3 (b), it is clearer that the shapes of the boundaries of the electronic phase regions are mixtures of a circle and a hexagon, and is more hexagonal than in Fig. 2 (b). The outlines of six-petal flowers in Fig. 3 (e) are also much easier to observe than in Fig. 2 (e). Furthermore, the mirror symmetry about the binary axis, mirror symmetry about the trigonal axis, and the associated inversion symmetry of the orientation-stoichiometry phase diagram of the trigonal-binary crystallographic plane normal to the bisectrix is further clarified in Fig. 3(c) and (f).

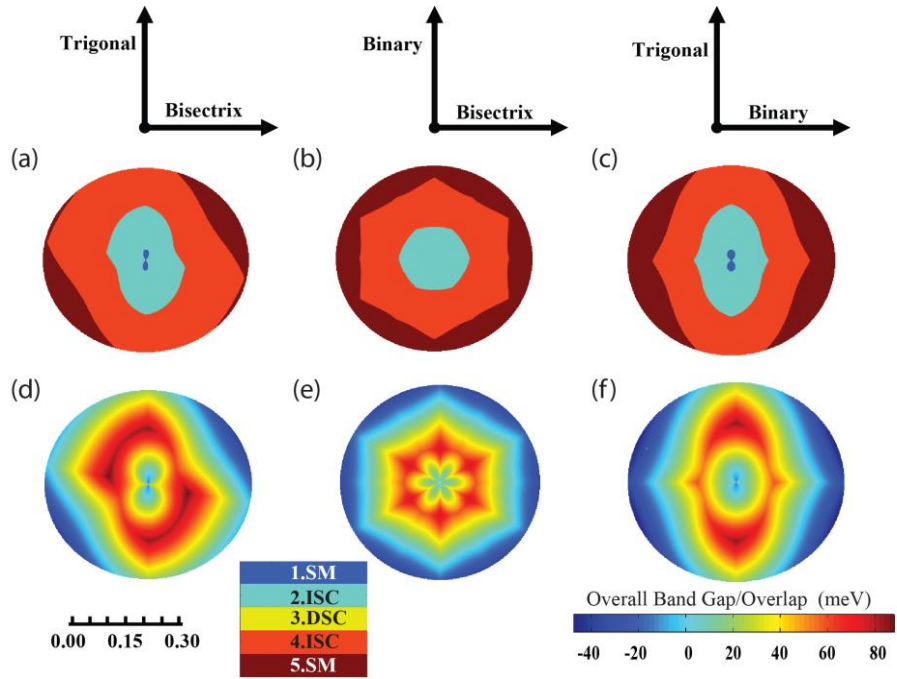

Figure 3: The electronic phase diagrams (a)-(c), and the band gap/overlap diagrams (d)-(f) of $Bi_{1-x}Sb_x$ nanowires of 10 nm wire width, as a function of wire growth orientation and Sb composition x. All the notations and legends are the same as those defined in Fig. 2.

The above discussions show that $Bi_{1-x}Sb_x$ nanowires of larger wire width show a much richer variation of electronic phases, but the contrast of anisotropy for different growth orientation is less obvious, while for the $Bi_{1-x}Sb_x$ nanowires of larger wire width, the richness of the variation of electronic phases is reduced, but the contrast of the anisotropy for different growth orientations is much enhanced. In order to show how the electronic and symmetry properties change with wire width and growth orientation, we have calculated the band gap/overlap diagrams as a function of wire width and growth orientation for both a small value of Sb composition of (x=0.05) and medium value of Sb composition (x=0.13), as shown in Fig. 4. The case of large values of Sb compositions turn out to be similar with the case of small values, only the location of the top of the valence band edges for the semimetal phases and the indirect semiconductor phases are at the *H* points for the median value of x=0.13, instead of at the *T* point which applies to (x=0.05). In order to make the values of band gap/overlap comparable in one plot, we choose to use a logarithm scale, and in order to avoid the divergent values of band shift when the wire width gets close to 0, we choose to ignore the cases where the wire width is smaller than $10^{0.5}{\approx}3$ nm.

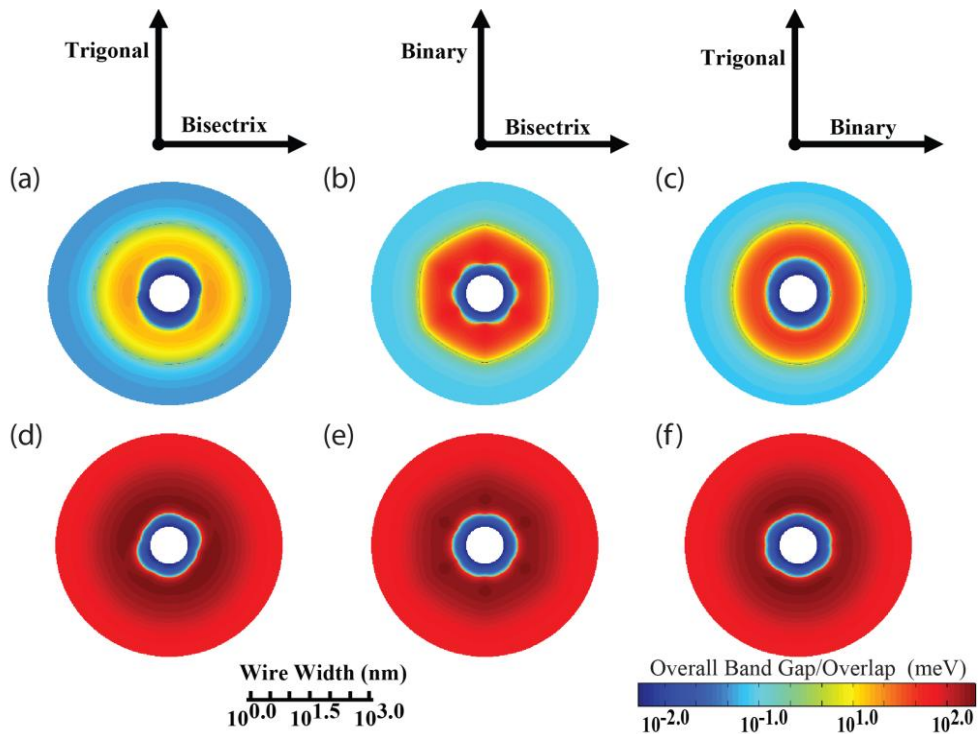

Figure 4: The band gap/overlap diagrams of $Bi_{1-x}Sb_x$ nanowires of small Sb composition (x=0.05)

(a)-(c) and medium Sb composition (x=0.13) (d)-(f), as a function of wire growth orientation and wire width. All the notations and legends are the same as those defined in Fig. 2 (d)-(f), except that the length of the radius in Fig. 4 stands for the wire width scaled logrartihmically, which lines up with the origin of each circularly shaped diagram.

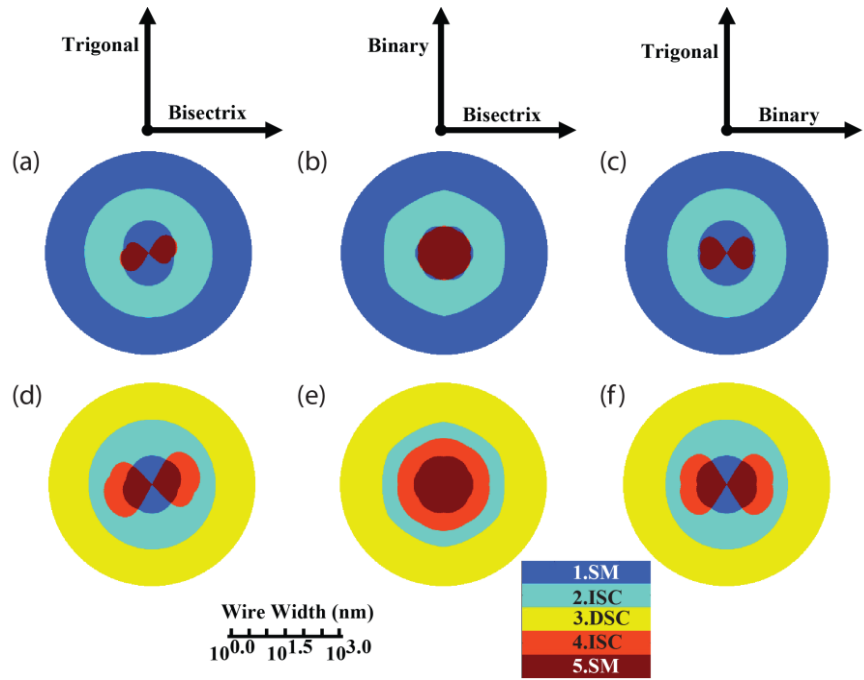

Figure 5: The electronic phase diagrams of $Bi_{1-x}Sb_x$ nanowires of small Sb composition (x=0.05) (a)-(c) and medium Sb composition (x=0.13) (d)-(f), as a function of wire growth orientation and wire width. All the notations and legends are the same with those defined in Fig. 2 (a)-(c), except for that the length of the radius in Fig. 4 stands for the wire width scaled logrartihmically.

For the small Sb composition (x=0.05) cases in Fig. 4 (a), (b) and (c), we see that there are mainly three regimes for each diagram: the inner semimetal regime when the wire width is small, the semiconductor regime when wire width is medium, and the outer semimetal regime when wire width is large. The outer semimetal regime is easy to understand, because the bulk bismuth antimony material at this Sb composition is in the semimetal phase region, with the top of the valence band edge located at the *T* point, as shown in Fig. 1c. The quantum confinement effect has induced an increase of the direct band gap region around the *L* points, and leads to a semiconductor phase at the medium wire width regime. That is also why the band overlap increases and saturates at a certain value with increasing wire width. However, the mechanism governing the inner semimetal regime is different. In our model, the components of the inverse-effective-mass tensor and the direct band gap at an *L* point are negatively correlated. When the width of a nanowire is large, the non-trivial quantum confinement effect of an *L* point shifts a band edge appear more quickly as the width decreases than does the trivial quantum confinement effect near an *H* or a *T* point. However, when the width of a nanowire is very small, the direct band gap at an *L* point is large enough to induce a significant reduction of the inter-band coupling, which changes the non-parabolic dispersions at the *L* point into parabolic dispersions with larger mass components. For this situation, the valence band edges at the *T* point and at the *H* points may, in contrast, have a larger quantum confinement effect. For the medium Sb composition ($x=0.13$) cases in Fig. 4 (d), (e) and (f), except for a small area near the center, where the wire width is very small, each band gap/overlap diagram has a positive value, and increases and saturates to a certain value with increasing wire width. This is also because the bulk bismuth-antimony material with this Sb composition is in the indirect semiconductor phase region, as shown in Fig. 1c. The corresponding phase diagrams as a function of wire width and growth orientation, for the small and medium Sb composition cases ($x=0.05$ and $x=0.13$) are shown in Fig. 5, which further solidifies the analysis above.

## IV.   CONCLUSION

In conclusion, we have developed a model to accurately describe the quantum confinement in one-dimensional narrow band gap systems, which accurately captures the band edge shifts and the shape of the non-parabolic dispersion relations. We have then used this model to study the phase diagrams and band gap/overlap of the bismuth-antimony nanowire materials system. The symmetry properties, the band edge configurations, the electronic phase diagrams, and the band gap/overlap diagrams as a function of wire growth orientation, stoichiomety and wire width are systematically studied. Such information provided an important guidance for the synthesis and characterization of bismuth antimony nanowires and for their use in the design of applications.

ACKNOWLEDGEMENT

The authors acknowledge the support from AFOSR MURI Grant number FA9550-10-1-0533, subaward 60028687. The views expressed are not endorsed by the sponsor.